\documentclass[aps,showpacs,twocolumn,prl]{revtex4}
\usepackage{colordvi}
\usepackage{amssymb}
\usepackage{amsmath}
\usepackage{mathrsfs}
\usepackage{mathlett}
\usepackage{graphicx,color,psfrag}
\begin{document}
\def \beq{\begin{equation}}
\def \eeq{\end{equation}}
\def \beqs{\begin{equation*}}
\def \eeqs{\end{equation*}}
\def \bea{\begin{eqnarray}}
\def \eea{\end{eqnarray}}
\def \bem{\begin{displaymath}}
\def \eem{\end{displaymath}}
\def \P{\Psi}
\def \Pd{|\Psi(\boldsymbol{r})|}
\def \Pds{|\Psi^{\ast}(\boldsymbol{r})|}
\def \Po{\overline{\Psi}}
\def \bs{\boldsymbol}
\def \bl{\bar{\boldsymbol{l}}}
\title{Electron optics with magnetic vector potential barriers in graphene}
\author{Sankalpa Ghosh and Manish Sharma}
\affiliation{Department of Physics, Indian Institute of Technology, Delhi, New Delhi-110016}
\begin{abstract}
An analysis of electron transport in graphene is  presented in the presence of various arrangement of 
delta-function like magnetic barriers. The motion through one such barrier gives an unusual non specular 
refraction leading to asymmetric transmission. The symmetry is restored by putting two such barriers in opposite
direction side by side. Periodic arrangements of such barriers can be used as  Bragg reflectors 
whose reflectivity has been calculated using a transfer matrix formalism. Such Bragg reflectors can be used to make 
resonant cavities. We also analyze the associated band structure for the case of infinite periodic structures.
\end{abstract}
\date{\today} 
\maketitle

In  a two-dimensional electron gas, there is a well established similarity 
between ballistic electron transport through electrostatic potential barriers 
and light propagation in geometrical optics \cite{SIVAN90}. This has been extended to the 
massless Dirac fermions in Graphene \cite{GeimreviewKSN1KSN2YZ1Castroneto}
where it was recently established that  
electron transport in the presence of an electrostatic potential barrier is analogous 
to negative refraction through metamaterials\cite{Falko, Veselago}. The relativistic behavior of Graphene electrons also leads to  Klein tunneling \cite{KSN3, Beenakker}, where a relativistic particle can tunnel through a high barrier by the process of pair production and thus it is not possible to confine them using such potential barriers. Such confinement is however possible by using magnetic barriers \cite{MDE07}. Can we understand this difference in behavior 
of massless Dirac fermions by comparing to propagation of light? The difficulty in using an optical analogy is that,
unlike the electrostatic potential, the magnetic vector potential couples with the momentum of the electron. 

In this work, we show that  wave-vector dependent tunneling of massless Dirac fermions 
 through magnetic barriers \cite{MPV94, PG95, ZC08, masir} can be understood in terms of well-known ideas in geometrical optics. However, the corresponding Snell's laws are very different from those of ordinary geometrical optics. We then carry out this analysis to propose devices such as a Bragg reflector using a transfer matrix approach and qualitatively depict how 
a resonant cavity can be constructed with such a reflector. Additionally, we
comment on the band structure of electron transport when such magnetic barriers 
are placed periodically.

\begin{figure}[ht]
\begin{center}
\resizebox{!}{5.25cm}{\includegraphics{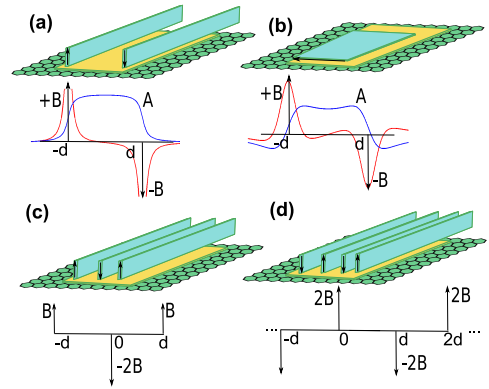}}%
\end{center}
\caption{\em
Monolayer graphene with ferromagnetic stripes having magnetizations perpendicular (a) and parallel (b) to plane. The magnetic field B (red) produces a magnetic vector potential A (blue). Single MVP barriers are formed in (a) and (b). Also shown are a double MVP barrier (c) and a periodic lattice (d).  }
\label{fig1}
\end{figure}

The proposed structure consists of a graphene sheet placed in close proximity to long magnetic stripes that produce highly localized magnetic fields as depicted in Fig.\ref{fig1}. Such field profiles can be created using demagnetizing fields produced at the edges of narrow stripes made with hard ferromagnetic materials of perpendicular or in-plane anisotropy. It is possible to make such stripes at various length scales. Materials such as CoCrPt used in magnetic recording produce field strengths of 1 Tesla close to the surface with bit lengths ranging from 50-100nm. Patterned stripes down to 10nm can be realized using nanolithography \cite{terriskrawczykbader}. It is possible to achieve field profiles at even smaller dimensions using domain walls with widths in the range of 10-50nm and magnetic nanostructures down to 5nm \cite{sunsci, zhengFePt} and 0.15nm \cite{bergmann} having highly localized field variations. 
The particular form of magnetic barrier that we shall use in our calculation can be realized using two narrow ferromagnetic stripes of perpendicular anisotropy with appropriately narrow dimensions and magnetized in opposite directions (Fig.\ref{fig1}(a)). The same profile can also be achieved by one ferromagnetic stripe whose magnetization is  parallel to the graphene sheet at a height $z_0$ above it (Fig.\ref{fig1}(b)). Such barriers have been used in the literature \cite{MPV94,PG95} and  the magnetic field of such structures is  
\beqs \bs{B}=B(x,z_0)\hat{z}=B_0[K(x+d,z_0)-K(x-d,z_0)]\hat{z} \eeqs 
Here, $K(x,z_0)=-\frac{2 z_0 d}{x^2 + z_0^2}$ and $B_0$ is a constant dependent on the aspect ratio of the stripe.  
For a given value of $z_0$ we plot in Fig.\ref{fig1}(a) the profile of such a magnetic field and the corresponding vector potential. As the plot shows, such an inhomogenous magnetic field can be well approximated  as a delta function-like magnetic barrier. This approximation is valid for all the length scales discussed in the preceding paragraph as long as the typical magnetic length $\ell_{B} = \sqrt{\frac{\hbar c}{|e|B}}$ is much larger than the width of such magnetic barriers. Accordingly, we use such delta function-like barriers for the rest of the paper. This choice is guided by the fact that it is amenable to simpler mathematical treatment, thus making  the analogy more transparent. We shall, however, point out later that the discussed analogy with geometrical optics is more general and is applicable for other types of magnetic barriers with finite width \cite{MDE07}. 

\begin{figure}[ht]
\begin{center}
\resizebox{!}{5.25cm}{\includegraphics{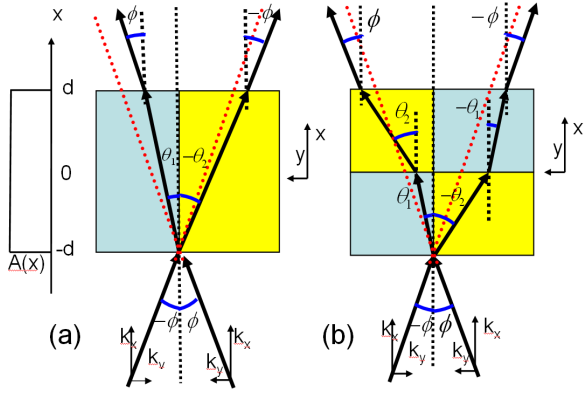}}%
\end{center}
\caption{\em (Color online) (a) Asymmetric refraction through a single barrier. (b) Refraction through a double barrier where the symmetry in the trasmission is restored. Barrier regions which are denser or rarer in terms of refraction of the wave vector are shaded with the same color.}
\label{fig8}
\end{figure}

We use the following magnetic field and vector potential in Landau gauge for a magnetic potential barrier \cite{MPV94}:
\beqs \bs{B}=B \ell_{B}[\delta(x+d)-\delta(x-d)]\hat{z}; \bs{A}_y(x) = B \ell_B \Theta(d^2-x^2)\hat{y} \eeqs
Since $\Theta(x)$ is the Heaviside step function,  we call this a magnetic vector potential (MVP) barrier.

For massless Dirac fermions in graphene in MVP barriers, 
we consider the limit where the electrons at the  $K$ and $K'$ points are decoupled from each other \cite{Ando}. 
Near each such point, the wavefunction is given by a two-component spinor and satisfies the equation
\beq v_F(\pi_x  \pm i \pi_y)\psi_{2,1} = E \psi_{1,2} \label{Eq1}\eeq
Here, $v_F$ is the fermi velocity ($\approx c/300$) and $\bs{\pi} = \bs{p}+\frac{e}{c}\bs{A}$.
Using $\frac{\hbar v_F}{\ell_B}$ as the unit of energy such that $\epsilon = \frac{E\ell_{B}}{\hbar v_F}$, $\ell_{B}$ as the unit of the length such that $\overline{x}=\frac{{x}}{\ell_{B}}$,
$\text{sgn} (e) =-1$ and $\psi = \phi(x)e^{ik_y\,y}$ in the Landau gauge, we get
\bea -i[ \frac{\partial}{\partial \overline{x}}  \pm (k_y \ell_B - \Delta)]\phi_{2,1} & = & \epsilon \phi_{1,2} \label{momentum}\eea
Here, $\Delta = 1$ for $|x|<d$ and $= 0$ for $|x|>d$. The above two coupled equations can be decoupled easily and result  in a Schr{\"{o}}dinger like equation of the form 
\beqs [- \frac{\partial^2}{\partial \overline{x}^2} +(k_y \ell_B  - \Delta)^2] \phi_{1,2}=  \epsilon ^2  \phi_{1,2}. \eeqs
In the region $ -d < x < d $  electrons see a barrier of height $[k_y + \text{sgn}(e) \frac{1}{\ell_B}]^2$. 
The corresponding wavefunctions in any region of space can be written in terms of a linear superposition of forward and backward moving plane waves such that
\beq
\phi_1 = \left\{
\begin{array}{rl}
e^{ik_x x} + r e^{-ik_x x} &  x < -d\\
a e^{iq_x x} + be^{-i q_x x}  & |x|<d \\
t e^{i k_x x} & x > d
\end{array} \right.
\label{particle}\eeq
\beq \phi_2 = \left\{
\begin{array}{rl}
s[e^{i (k_x x +  \phi)} - r e^{-i(k_x x + \phi)}] & x < -d\\
s'[a e^{i(q_x x + \theta)} - b e^{-i(q_x x + \theta)}] &    |x|<d \\ 
s te^{i(k_x x +\phi)} &  x > d
\end{array} \right.
\label{hole}\eeq
Solutions of the above equations are very different from those in the presence of a uniform magnetic field 
since here the magnetic field is highly non-uniform and has singular delta function like structures. Also, $s,s'$ are given by $\text{sgn}(\epsilon)$ and are both $+1$ for electrons when only magnetic fields are present and no electrostatic potentials are applied. A similar treatment can be done in the presence of an additional electrostatic potential, in which case both $s,s'$ are not necessarily $+1$ \cite{MSSG2}. The wave vector components are $[k_x , k_y] = k_F [cos \phi, sin \phi]$ outside the magnetic barrier and $\phi$ is the incident angle for an electron wave.
The fermi energy of the incident electrons is $E_F = \hbar v_F k_F$. In the dimensionless form, this is $\epsilon_F = k_F \ell_B$ and will control the transport by changing the refractive index of the barrier region. Since magnetic field does not do any work, energy conservation gives $k_x^2+k_y^2=k_F^2$ for ${|x|>d}$ and $q_x^2+(k_y-\frac{1}{\ell_B})^2=k_F^2$ for $|x|<d$.
Since $\theta = cos^{-1}(\frac{q_{x}}{k_F})$, for $|x|<d$
$k_F \sin \theta = k_y - \frac{1}{\ell_B}$ gives
\beq \sin |\theta| = \sin |\phi| -\text{sgn}(\phi)\frac{1}{k_F \ell_B}, -\frac{\pi}{2} < \phi < \frac{\pi}{2} 
\label{Snells1} \eeq 
The situation has  been depicted in Fig. \ref{fig8}(a).
The wave incident with positive $\phi$ wave vector will bend towards the normal whereas the waves incident with 
negative incidence angle, corresponding wave vector will bend away from the surface normal inside the barrier
region.  
Therefore, the Snell's law of electron waves in such magnetic barriers 
is not specular as it is for light 
wave on smooth surface or  for the incidence of the electrons on an 
electrostatic potential barrier \cite{Falko, KSN3}. This unusual refraction  
can also be thought as a consequence of breaking of time reversal symmetry in presence of such magnetic barriers. When the magnetic field will be reversed, the denser and rarer medium will interchange their side without changing the asymmtric transmission. 

\begin{figure}[ht]
\begin{center}
\begin{psfrags}%
\psfragscanon%
%
\psfrag{s05}[b][b]{\color[rgb]{0,0,0}\setlength{\tabcolsep}{0pt}\begin{tabular}{c}{\bf (a)}\end{tabular}}%
\psfrag{s06}[][]{\color[rgb]{0,0,0}\setlength{\tabcolsep}{0pt}\begin{tabular}{c}$0.5$\end{tabular}}%
\psfrag{s07}[][]{\color[rgb]{0,0,0}\setlength{\tabcolsep}{0pt}\begin{tabular}{c}$1$\end{tabular}}%
\psfrag{s08}[][]{\color[rgb]{0,0,0}\setlength{\tabcolsep}{0pt}\begin{tabular}{c}$-{\pi}/{2}$\end{tabular}}%
\psfrag{s09}[][]{\color[rgb]{0,0,0}\setlength{\tabcolsep}{0pt}\begin{tabular}{c}$-{\pi}/{4}$\end{tabular}}%
\psfrag{s10}[][]{\color[rgb]{0,0,0}\setlength{\tabcolsep}{0pt}\begin{tabular}{c}${\pi}/{4}$\end{tabular}}%
\psfrag{s11}[][]{\color[rgb]{0,0,0}\setlength{\tabcolsep}{0pt}\begin{tabular}{c}${\pi}/{2}$\end{tabular}}%
\psfrag{s14}[b][b]{\color[rgb]{0,0,0}\setlength{\tabcolsep}{0pt}\begin{tabular}{c}{\bf (b)}\end{tabular}}%
\psfrag{s15}[][]{\color[rgb]{0,0,0}\setlength{\tabcolsep}{0pt}\begin{tabular}{c}$0.5$\end{tabular}}%
\psfrag{s16}[][]{\color[rgb]{0,0,0}\setlength{\tabcolsep}{0pt}\begin{tabular}{c}$1$\end{tabular}}%
\psfrag{s17}[][]{\color[rgb]{0,0,0}\setlength{\tabcolsep}{0pt}\begin{tabular}{c}$-{\pi}/{2}$\end{tabular}}%
\psfrag{s18}[][]{\color[rgb]{0,0,0}\setlength{\tabcolsep}{0pt}\begin{tabular}{c}$-{\pi}/{4}$\end{tabular}}%
\psfrag{s19}[][]{\color[rgb]{0,0,0}\setlength{\tabcolsep}{0pt}\begin{tabular}{c}${\pi}/{4}$\end{tabular}}%
\psfrag{s20}[][]{\color[rgb]{0,0,0}\setlength{\tabcolsep}{0pt}\begin{tabular}{c}${\pi}/{2}$\end{tabular}}%
%
\psfrag{x01}[t][t]{$0$}%
\psfrag{x02}[t][t]{$0.2$}%
\psfrag{x03}[t][t]{$0.4$}%
\psfrag{x04}[t][t]{$0.6$}%
\psfrag{x05}[t][t]{$0.8$}%
\psfrag{x06}[t][t]{$1$}%
\psfrag{x07}[t][t]{$0$}%
\psfrag{x08}[t][t]{$0.2$}%
\psfrag{x09}[t][t]{$0.4$}%
\psfrag{x10}[t][t]{$0.6$}%
\psfrag{x11}[t][t]{$0.8$}%
\psfrag{x12}[t][t]{$1$}%
%
\psfrag{v01}[r][r]{$-1$}%
\psfrag{v02}[r][r]{$-0.8$}%
\psfrag{v03}[r][r]{$-0.6$}%
\psfrag{v04}[r][r]{$-0.4$}%
\psfrag{v05}[r][r]{$-0.2$}%
\psfrag{v06}[r][r]{$0$}%
\psfrag{v07}[r][r]{$0.2$}%
\psfrag{v08}[r][r]{$0.4$}%
\psfrag{v09}[r][r]{$0.6$}%
\psfrag{v10}[r][r]{$0.8$}%
\psfrag{v11}[r][r]{$1$}%
\psfrag{v12}[r][r]{$-1$}%
\psfrag{v13}[r][r]{$-0.8$}%
\psfrag{v14}[r][r]{$-0.6$}%
\psfrag{v15}[r][r]{$-0.4$}%
\psfrag{v16}[r][r]{$-0.2$}%
\psfrag{v17}[r][r]{$0$}%
\psfrag{v18}[r][r]{$0.2$}%
\psfrag{v19}[r][r]{$0.4$}%
\psfrag{v20}[r][r]{$0.6$}%
\psfrag{v21}[r][r]{$0.8$}%
\psfrag{v22}[r][r]{$1$}%
%
\resizebox{!}{5cm}{\includegraphics{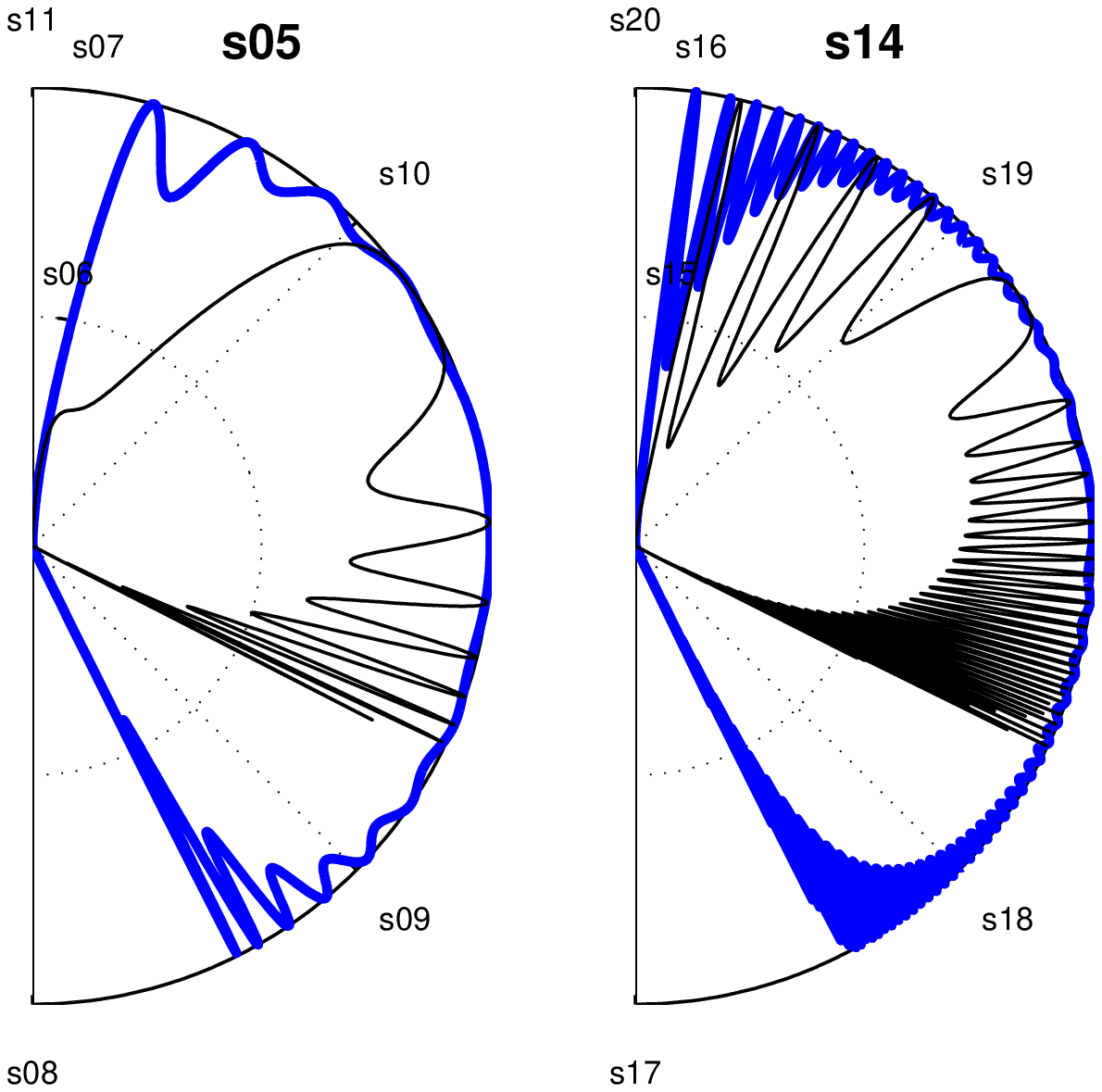}}%
\end{psfrags}
\end{center}
\caption{\em
Polar plot of T vs. $\phi$ for single MVP barriers of widths (a) 100nm, (b) 500nm. Blue (thick)=$0.1$T with $l_B$=$81$nm, Black (thin)=$3$T with $l_B$=$14.8$nm.}
\label{fig2a}
\end{figure}

According to Eq. \ref{Snells1}, when $|\sin |\theta|| > 1$, $\theta$ becomes 
imaginary and the wave in the second medium becomes evanescent. In the language 
of optics this corresponds to total internal reflection (TIR). According  to Fig. \ref{fig8}(a), this will happen when $\sin |\theta| > 1$
for  ${-\frac{\pi}{2}} \le \phi < 0$ and  when $\sin |\theta| < -1$ 
for  $0 < \phi 
\le  \frac{\pi}{2}$. In the latter case, this requires the wave vector to be negatively refracted \cite{Falko} at sufficiently high magnetic field before 
TIR occurs. It also follows that for a given strength $B$
 the magnitude of critical angle of 
incidence $|\phi|=\phi_c$ for TIR is  higher for $0 < \phi < \frac{\pi}{2}$
as compared to the one for 
$ -\frac{\pi}{2} \le \phi < 0 $. Because of TIR the transmission on both 
sides of Fig \ref{fig2a} drops to $0$ beyond a certain value of $\phi$ and 
this value is lower for negative angles of incidence . 

The wavefunctions given in Eqs. \ref{particle} and \ref{hole} are  similar to those of  massless Dirac electrons 
scattered by an electrostatic  step potential considered in \cite{KSN3}.
This is because the MVP barrier creates a momentum dependent step potential of $[k_y + \text{sgn}(e) \frac{1}{\ell_B}]^2 $.
Continuity of the wavefunction at the boundaries of the MVP barrier can be used to calculate the transmission coefficient as
\beq t = \frac{2ss'e^{-ik_{x}D}\cos\phi\cos\theta}{ss'[e^{-iq_xD}\cos(\phi+\theta)+e^{iq_x D}\cos(\phi-\theta)]-2i\sin{q_xD}} \label{transmission} \eeq
Here, $D=2d$. Thus  $t$, transmittance $T=t^{*}{t}$ and reflectance  $R=1-T$ have same expressions 
for electrostatic potentials in Ref.\cite{KSN3}. There are, however, key differences. 
For electrostatic barrier as $\phi \rightarrow -\phi$, $\theta \rightarrow -\theta$. Here because of Eq.\ref{Snells1} that is not the case. Thus, the same  Eq.\ref{transmission} gives symmetric transmission for electrostatic potential in \cite{KSN3} and asymmetric transmission here. 

For high electrostatic barriers such that $V \gg E_{F}$, the wave vector is given by $q_x=\sqrt{\frac{(E_{F}-V)^2}{\hbar^2 v_{F}^2}-k_y^2}$, which is real. The corresponding transmittance for electrostatic potentials is 
\beq T = \frac{\cos ^2 \phi}{1 - \cos^(q_x D) \sin^2 \phi} \label{Klein} \eeq
and  is $1$ at $\phi=0$. This exhibits  Klein tunneling for massless Dirac fermions \cite{KSN3}.
For the magnetic barrier, $\frac{1}{k_F \ell_B} \propto \sqrt{B}$. However, unlike
the electrostatic field case, the magnetic field changes the 
wave vector in both  direction and not the energy. At high magnetic field, $q_x^2 = k_F^2 - (k_y - \frac{1}{\ell_B})^2 = -\kappa^2 < 0$.  As discussed, this leads to TIR and not Klein tunneling. In  Fig. \ref{fig2a},
the magnitude of critical angle beyond which TIR occurs is lower for a higher magnetic field.  Then, a stronger  MVP barrier leads to higher reflections  as opposed to complete transmission at normal incidence 
by  high electrostatic potential barrier. A similar situation is encountered in other forms of magnetic barriers \cite{MDE07, GhoshOroszlany}.

Complete transmission only occurs for $q_xD = n\pi$ in  Eq.\ref{transmission}. This corresponds to resonant tunneling for Dirac electrons and happens in the same way as for non-relativistic electrons, appearing as a number of peaks in the plots in Fig.\ref{fig2a}. The number of such tunneling peaks increases with barrier width for both MVP barriers and electrostatic barriers.

\begin{figure}[ht]
\begin{center}
\begin{psfrags}%
\psfragscanon%
%
\psfrag{s05}[b][b]{\color[rgb]{0,0,0}\setlength{\tabcolsep}{0pt}\begin{tabular}{c}{\bf (a)}\end{tabular}}%
\psfrag{s06}[][]{\color[rgb]{0,0,0}\setlength{\tabcolsep}{0pt}\begin{tabular}{c}$0.5$\end{tabular}}%
\psfrag{s07}[][]{\color[rgb]{0,0,0}\setlength{\tabcolsep}{0pt}\begin{tabular}{c}$1$\end{tabular}}%
\psfrag{s08}[][]{\color[rgb]{0,0,0}\setlength{\tabcolsep}{0pt}\begin{tabular}{c}$-{\pi}/{2}$\end{tabular}}%
\psfrag{s09}[][]{\color[rgb]{0,0,0}\setlength{\tabcolsep}{0pt}\begin{tabular}{c}$-{\pi}/{4}$\end{tabular}}%
\psfrag{s10}[][]{\color[rgb]{0,0,0}\setlength{\tabcolsep}{0pt}\begin{tabular}{c}${\pi}/{4}$\end{tabular}}%
\psfrag{s11}[][]{\color[rgb]{0,0,0}\setlength{\tabcolsep}{0pt}\begin{tabular}{c}${\pi}/{2}$\end{tabular}}%
\psfrag{s14}[b][b]{\color[rgb]{0,0,0}\setlength{\tabcolsep}{0pt}\begin{tabular}{c}{\bf (b)}\end{tabular}}%
\psfrag{s15}[][]{\color[rgb]{0,0,0}\setlength{\tabcolsep}{0pt}\begin{tabular}{c}$0.5$\end{tabular}}%
\psfrag{s16}[][]{\color[rgb]{0,0,0}\setlength{\tabcolsep}{0pt}\begin{tabular}{c}$1$\end{tabular}}%
\psfrag{s17}[][]{\color[rgb]{0,0,0}\setlength{\tabcolsep}{0pt}\begin{tabular}{c}$-{\pi}/{2}$\end{tabular}}%
\psfrag{s18}[][]{\color[rgb]{0,0,0}\setlength{\tabcolsep}{0pt}\begin{tabular}{c}$-{\pi}/{4}$\end{tabular}}%
\psfrag{s19}[][]{\color[rgb]{0,0,0}\setlength{\tabcolsep}{0pt}\begin{tabular}{c}${\pi}/{4}$\end{tabular}}%
\psfrag{s20}[][]{\color[rgb]{0,0,0}\setlength{\tabcolsep}{0pt}\begin{tabular}{c}${\pi}/{2}$\end{tabular}}%
%
\psfrag{x01}[t][t]{$0$}%
\psfrag{x02}[t][t]{$0.2$}%
\psfrag{x03}[t][t]{$0.4$}%
\psfrag{x04}[t][t]{$0.6$}%
\psfrag{x05}[t][t]{$0.8$}%
\psfrag{x06}[t][t]{$1$}%
\psfrag{x07}[t][t]{$0$}%
\psfrag{x08}[t][t]{$0.2$}%
\psfrag{x09}[t][t]{$0.4$}%
\psfrag{x10}[t][t]{$0.6$}%
\psfrag{x11}[t][t]{$0.8$}%
\psfrag{x12}[t][t]{$1$}%
%
\psfrag{v01}[r][r]{$-1$}%
\psfrag{v02}[r][r]{$-0.8$}%
\psfrag{v03}[r][r]{$-0.6$}%
\psfrag{v04}[r][r]{$-0.4$}%
\psfrag{v05}[r][r]{$-0.2$}%
\psfrag{v06}[r][r]{$0$}%
\psfrag{v07}[r][r]{$0.2$}%
\psfrag{v08}[r][r]{$0.4$}%
\psfrag{v09}[r][r]{$0.6$}%
\psfrag{v10}[r][r]{$0.8$}%
\psfrag{v11}[r][r]{$1$}%
\psfrag{v12}[r][r]{$-1$}%
\psfrag{v13}[r][r]{$-0.8$}%
\psfrag{v14}[r][r]{$-0.6$}%
\psfrag{v15}[r][r]{$-0.4$}%
\psfrag{v16}[r][r]{$-0.2$}%
\psfrag{v17}[r][r]{$0$}%
\psfrag{v18}[r][r]{$0.2$}%
\psfrag{v19}[r][r]{$0.4$}%
\psfrag{v20}[r][r]{$0.6$}%
\psfrag{v21}[r][r]{$0.8$}%
\psfrag{v22}[r][r]{$1$}%
%
\resizebox{!}{5cm}{\includegraphics{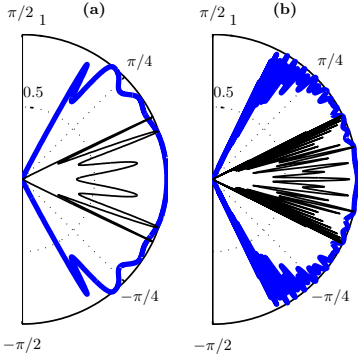}}%
\end{psfrags}%
\end{center}
\caption{\em
Polar plot of T vs. $\phi$ for DMVP barriers of widths (a) 100nm, (b) 500nm. Blue (thick)=$0.1$T with $l_B$=$81$nm, Black (thin)=3T with $l_B$=$14.8$nm.}
\label{fig2b}
\end{figure}

To get symmetric transmission out of such a barrier it is therefore required
to place two such MVP barriers side by side but oppositely oriented as depicted in  
 Fig.\ref{fig1}(c). The magnetic field creating such a barrier is
\beqs \bs{B}=B_z(x)\hat{z}=B\ell_B[\delta(x+d)+ \delta(x-d) -2 B \delta(x)]\hat{z}  \eeqs
We again consider energy conservation in medium $1$ ($-d < x < 0$) and medium $2$ ($0 < x < d$), which gives
\beq q_{1,2}^2 + (k_y \mp \frac{1}{\ell_B})^2  =  k_F^2; \sin |\theta_{1,2}| 
=  \sin |\phi| \mp \text{sgn}(\phi)\frac{1}{k_F \ell_B } \eeq
The incident angle is $-\frac{\pi}{2}<\phi<\frac{\pi}{2}$ and the angle of refraction is $\theta_1$ and $\theta_2$ in media $1$ and $2$ respectively.
The absolute value of relative refractive index  of region $1$ with respect to region $2$ 
on the left side of the surface normal is just the inverse of that on the right
side of the surface normal and can be combined in the following expression 
\beq |_{1}n_{2}|=\frac{\sin|\theta_1|}{\sin |\theta_2|} = 
\frac{\sin |\phi| - \text{sgn}(\phi)\frac{1}{k_F \ell_B}}{\sin |\phi| + 
\text{sgn}(\phi)\frac{1}{k_F \ell_B}} \label{rindex} \eeq
Thus, for such double MVP barriers, the 
wave vector bending towards (away from)
the surface normal in the first half of the barrier bends away from (towards) the surface normal in the second half of the barrier as shown in Fig.\ref{fig8}(b). 
This will achieve symmetric transmission through such a barrier as demonstrated 
in Fig.\ref{fig2b}. Critical angles of incidence beyond which TIR will occur
for positive and negative $\phi$ will also be interchanged as we go from the first
to the second barrier region. However, at higher $B$ fields, TIR will occur at both 
regions of the barrier. Consequently, the total reflectivity of the 
barrier increases as can be seen by comparing Figs.\ref{fig2a} and \ref{fig2b}.

Practical devices such as Bragg reflectors 
 can be made by exploiting the high reflectivity of 
double MVP barriers to manipulate electrons, as will be discussed later.
Such structures, if realized, could also be very useful to confine carriers in desired areas in graphene and away from edges, where edge states could adversely affect transport. 

To calculate transmittance for double MVP (DMVP) barriers, we write the wavefunction in the same way as in Eqs.\ref{particle} and \ref{hole}:
\beq
\phi_1 = \left\{
\begin{array}{rl}
e^{ik_x x} + r e^{-ik_x x} &  x < -d\\
a e^{iq_1 x} + be^{-i q_1 x}  & x\in[-d,0] \\
c e^{iq_2 x} + de^{-i q_2 x}  & x\in[0,d] \\
t e^{i k_x x} & x > d
\end{array} \right.
\label{particle2}\eeq
\beq \phi_2 = \left\{
\begin{array}{rl}
s[e^{i (k_x x +  \phi)} - r e^{-i(k_x x + \phi)}] & x < -d\\
s_1[a e^{i(q_1 x + \theta_1)} - b e^{-i(q_1 x + \theta_1)}] &   x \in[-d,0]\\
s_2[c e^{i(q_2 x + \theta_2)} - d e^{-i(q_2 x + \theta_2)}] &    x\in[0,d]\\
s te^{i(k_x x +\phi)} & x > d
\end{array} \right.
\label{hole2}\eeq
The corresponding sign factors associated with $\phi_2$ in these regions are $s_1$ and $s_2$ and are both $1$. 
The transmittance and reflectance can now be easily computed by imposing the continuity conditions on the above functions at the location of the barriers, namely at $x=-d,0,d$. We can express the transmittance and reflectance in a compact form by introducing
 $A=e^{-ik_x x}$ and $B_{1,2}= e^{-iq_{1,2}x}$ to define the following matrices:
\bea M_A = \begin{bmatrix}A & 0 \\ 0 & A^*\end{bmatrix},
M_{\theta_{1,2},\phi} = \begin{bmatrix}1 & 1 \\ e^{i\theta_{1,2},\phi} & -e^{-i\theta_{1,2},\phi}\end{bmatrix} \nonumber \\
M_{s,s_1,s_2} = \begin{bmatrix} 1 & 0 \\ 0 & s,s_1,s_2 \end{bmatrix} = I,
M_{B_{1,2}} = \begin{bmatrix}B_{1,2} & 0 \\ 0 & B_{1,2}^*\end{bmatrix}  \nonumber
\eea
Here $I$ is the unit matrix and 
\beq 
M_{A}^{-1}=M_{A}^{*}; M_{B_{12}}^{-1} = M_{B_{12}}^{*}
\label{aastar}
\eeq
The solution of these continuity equations can then be written as 
\bea \begin{bmatrix}1  & r \end{bmatrix}^{T} & = & M_A^{*}{M_\phi}^{-1}[M_{\theta_1}M_{B_1}M_{\theta_1}^{-1} \nonumber \\
&  & M_{\theta_2}M_{B_2}M_{\theta_2}^{-1}]M_{\phi}M_{A^*}\begin{bmatrix}t & 0\end{bmatrix}^{T} \nonumber \eea
To understand the above formulae, we introduce the transfer matrix through a DMVP barrier:
\beqs T_\text{DMVP} = M_{\theta_1} M_{B_1} M_{\theta_1}^{-1}M_{\theta_2} M_{B_2}M_{\theta_2}^{-1}  \eeqs
We can interpret $M_{\phi}^{-1}M_{\theta_1}$ as the phase shift at the first boundary, $M_{\theta_2}^{-1} M_{\phi}$ as the phase shift at the last boundary
and $M_{\theta_1}^{-1}M_{\theta_2}$ as the phase shift at the barrier at $x=0$. 
Thus, we can rewrite the above equation as 
\beq \begin{bmatrix}1  & r \end{bmatrix}^{T}  =  M_A^{*}{M_\phi}^{-1}T_\text{DMVP}M_{\phi}M_{A^*}\begin{bmatrix}t & 0\end{bmatrix}^{T} \label{DMVP} \eeq
The resulting two equations can be  solved to yield the transmittance through DMVP barriers, which has been plotted in Fig.\ref{fig2b}. 

\begin{figure}[ht]
\begin{center}
\begin{psfrags}%
\psfragscanon%
%
\psfrag{s10}[t][t]{\color[rgb]{0,0,0}\setlength{\tabcolsep}{0pt}\begin{tabular}{c}B=0.1T\end{tabular}}%
\psfrag{s11}[t][t]{\color[rgb]{0,0,0}\setlength{\tabcolsep}{0pt}\begin{tabular}{c}B=1T\end{tabular}}%
\psfrag{s12}[t][t]{\color[rgb]{0,0,0}\setlength{\tabcolsep}{0pt}\begin{tabular}{c}B=3T\end{tabular}}%
%
\psfrag{x01}[t][t]{$-\pi/2$}%
\psfrag{x02}[t][t]{$-\pi/4$}%
\psfrag{x03}[t][t]{$0$}%
\psfrag{x04}[t][t]{$\pi/4$}%
\psfrag{x05}[t][t]{$\pi/2$}%
\psfrag{x06}[t][t]{$-1.5$}%
\psfrag{x07}[t][t]{$-1$}%
\psfrag{x08}[t][t]{$-0.5$}%
\psfrag{x09}[t][t]{$0$}%
\psfrag{x10}[t][t]{$0.5$}%
\psfrag{x11}[t][t]{$1$}%
\psfrag{x12}[t][t]{$1.5$}%
\psfrag{x13}[t][t]{$-1.5$}%
\psfrag{x14}[t][t]{$-1$}%
\psfrag{x15}[t][t]{$-0.5$}%
\psfrag{x16}[t][t]{$0$}%
\psfrag{x17}[t][t]{$0.5$}%
\psfrag{x18}[t][t]{$1$}%
\psfrag{x19}[t][t]{$1.5$}%
%
\psfrag{v01}[r][r]{$-1$}%
\psfrag{v02}[r][r]{$0$}%
\psfrag{v03}[r][r]{$1$}%
\psfrag{v04}[r][r]{$-1$}%
\psfrag{v05}[r][r]{$0$}%
\psfrag{v06}[r][r]{$1$}%
\psfrag{v07}[r][r]{$-1$}%
\psfrag{v08}[r][r]{$0$}%
\psfrag{v09}[r][r]{$1$}%
%
\resizebox{7.8cm}{!}{\includegraphics{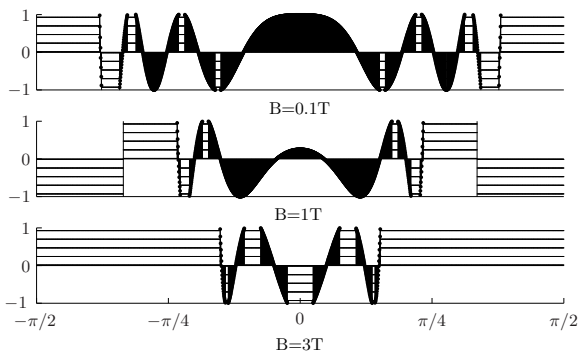}}%
\end{psfrags}%
%
\end{center}
\caption{{\em $\cos(KD)$ vs. $\phi$ for an infinite periodic lattice with $d$=$100$nm showing evanescent (dashed) and propagating (solid) Bloch waves. For comparison with earlier plots, $\phi$ is still used.}}
\label{fig4}
\end{figure}

We shall now discuss similarities between transport of massless Dirac fermions through MVP barriers with  electromagnetic propagation in periodic stratified media \cite{yariv1} as well as Dirac fermions in periodic electrostatic potentials \cite{periodic}. In real structures, there will only be a finite number of barriers and the lattice translational symmetry will break down at the boundary. To simplify the analysis, we assume the DMVP barrier structure can be repeated infinitely.
We consider each unit cell of size $D=2d$. The $n$-th cell is given by $(n-1)D < x < nD$.
In the $\alpha$-th part of the given unit cell, the wavefunction is
\bea \phi_1 & = & a_{n}^{\alpha} e^{i q^{n}_{\alpha x} (x - nD)} + b_{n}^{\alpha}e^{-i q^{n}_{\alpha x} (x - nD)} \nonumber \\
\phi_2 & = & s_{n}^{\alpha} \left[a_{n}^{\alpha} e^{i [q^{n}_{\alpha x} (x - nD)+ \theta_{\alpha}]} - b_{n}^{\alpha}e^{-i [q^{n}_{\alpha x} (x - nD)+ \theta_{\alpha}] } \right] \nonumber \eea 
Here, $\alpha  =  1,2, a_{n}^{1}  =  a_n, b_{n}^{1}  =  b_n, a_{n}^{2}  =  c_n, b_{n}^{2}  =  d_n, q^{n}_{1x} =  q_{1}, q^{n}_{2x} =  q_{2} $. The exponential factor $e^{-inD}$ reveals the lattice translational symmetry, which is not present for isolated single and double barrier structures.
Imposing continuity at both interfaces of the $n$-th unit cell gives
\bea M_{s_{{2},{n-1}}}M_{\theta_2}\begin{bmatrix} c_{n-1} \\
d_{n-1} \end{bmatrix} &= & M_{s_{1,n}}M_{\theta_1} {M_{B_1}}^2\begin{bmatrix} a_{n} \\
b_{n} \end{bmatrix} \nonumber \\
M_{s_{1,n}}M_{\theta_1}M_{B_1}\begin{bmatrix} a_{n} \\ b_{n} \end{bmatrix} & = &
M_{s_{2,n}}{M_{\theta_2}}{M_{B_2}}\begin{bmatrix} c_{n} \\ d_{n} \end{bmatrix} \label{latticebc}\eea
Imposing Bloch condition gives the band structure from the following eigenequations
\beqs \begin{bmatrix} c_{n} \\ d_{n} \end{bmatrix} = e^{iK D} \begin{bmatrix} c_{n-1} \\ d_{n-1} \end{bmatrix}= \begin{bmatrix} K_{11} & K_{12} \\
K_{21} & K_{22} \end{bmatrix} \begin{bmatrix} c_{n-1} \\ d_{n-1} \end{bmatrix} \eeqs
where $K$ is the Bloch momentum and the matrix elements $K_{ij}$ can be calculated  from Eq.\ref{latticebc} as
\beqs  \begin{bmatrix} K_{11} & K_{12} \\
K_{21} & K_{22} \end{bmatrix} = (M_{\theta_2}M_{B_2})^{-1}M_{\theta_1}(M_{\theta_1}M_{B_1})^{-1}M_{\theta_2}
\eeqs
This is the same as $T_\text{DMVP}^{-1}$ permuted. Unitarity gives $\det{K_{{ij}}}=1$, yielding the eigenvalue equation
\beqs K(\phi, B) = \frac{1}{2d} \cos^{-1}[\frac{1}{2}Tr(K_{ij})] \eeqs 
The condition $|\frac{1}{2}Tr(K_{{ij}})| < 1$ corresponds to propagating Bloch waves whereas $|\frac{1}{2}Tr(K_{ij})| > 1$ leads to evanescent Bloch waves that correspond to forbidden zones in the band structure. Such band structures have previously been studied for many problems including condensed matter systems, optics \cite{yariv1} and relativistic quarks
\cite{periodic}. While the detailed band structure is provided elsewhere \cite{MSSG2}, we plot in Fig.\ref{fig4}  $|\frac{1}{2}Tr(K_{ij})|$ as a function of the incident angle $\phi$  for different $B$ values. A forbidden region appears at $\phi=0$ at higher $B$ due to larger difference between the refractive indices of adjacent regions.

\begin{figure}[ht]
\begin{center}
\begin{psfrags}%
\psfragscanon%
%
\psfrag{s10}[t][t]{\color[rgb]{0,0,0}\setlength{\tabcolsep}{0pt}\begin{tabular}{c}n=2\end{tabular}}%
\psfrag{s11}[t][t]{\color[rgb]{0,0,0}\setlength{\tabcolsep}{0pt}\begin{tabular}{c}n=5\end{tabular}}%
\psfrag{s12}[t][t]{\color[rgb]{0,0,0}\setlength{\tabcolsep}{0pt}\begin{tabular}{c}n=25\end{tabular}}%
%
\psfrag{x01}[t][t]{$-\pi/2$}%
\psfrag{x02}[t][t]{$-\pi/4$}%
\psfrag{x03}[t][t]{$0$}%
\psfrag{x04}[t][t]{$\pi/4$}%
\psfrag{x05}[t][t]{$\pi/2$}%
\psfrag{x06}[t][t]{$-1.5$}%
\psfrag{x07}[t][t]{$-1$}%
\psfrag{x08}[t][t]{$-0.5$}%
\psfrag{x09}[t][t]{$0$}%
\psfrag{x10}[t][t]{$0.5$}%
\psfrag{x11}[t][t]{$1$}%
\psfrag{x12}[t][t]{$1.5$}%
\psfrag{x13}[t][t]{$-1.5$}%
\psfrag{x14}[t][t]{$-1$}%
\psfrag{x15}[t][t]{$-0.5$}%
\psfrag{x16}[t][t]{$0$}%
\psfrag{x17}[t][t]{$0.5$}%
\psfrag{x18}[t][t]{$1$}%
\psfrag{x19}[t][t]{$1.5$}%
%
\psfrag{v01}[r][r]{$0$}%
\psfrag{v02}[r][r]{$0.5$}%
\psfrag{v03}[r][r]{$1$}%
\psfrag{v04}[r][r]{$0$}%
\psfrag{v05}[r][r]{$0.5$}%
\psfrag{v06}[r][r]{$1$}%
\psfrag{v07}[r][r]{$0$}%
\psfrag{v08}[r][r]{$0.5$}%
\psfrag{v09}[r][r]{$1$}%
%
\resizebox{7.8cm}{!}{\includegraphics{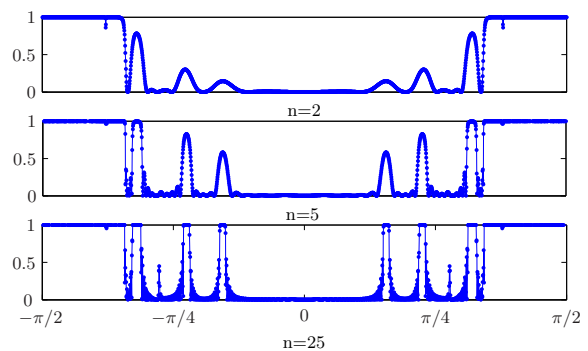}}%
\end{psfrags}%
\end{center}
\caption{\em Reflectance versus $\phi$ through a Bragg reflector with different periods of DMVP barriers of d=$100$nm and B=$0.1$T.}
\label{fig6}
\end{figure}

We shall now modify the above result of infinite periodic barriers to analyze a finite chain 
of DMVP barriers that  makes a Bragg reflector. A Bragg reflector can be formed by superposing side by side $n$ such DMVP barriers. We shall briefly describe how the transmittance and 
reflectance of such a reflector can be calculated using a transfer matrix formalism. 
The magnetic field for a Bragg reflector placed symmetrically around the origin can be written as 
\bea
\bs{B} = B_z(x)\hat{z} &=& B\ell_B[\delta(x+nd)+ \delta(x-nd) \nonumber \\
 & & + \sum_{p=1-n}^{n-1}(-1)^{p+n}2B\delta(x-p d))]\hat{z}   \eea
The series of wave function solutions in the various regions are linear combinations of right and left moving waves similar to the ones given earlier in Eqs. \ref{particle2} and \ref{hole2} for one DMVP barrier.
To solve this set of equations, we proceed as earlier using continuity of the wave function at the magnetic barriers at $x=p d, -n \le p \le n$. Just as Eq.\ref{DMVP} describes the solution for one DMVP barrier, the solutions for $n$ DMVP barriers can be written in matrix form as
\beq \begin{bmatrix}1  & r \end{bmatrix}^{T}  = ({M_A}^{-1})^{n}M_{\phi}^{-1} T_\text{DMVP}^n M_{\phi} (M_{A^{*}})^{n} 
\begin{bmatrix}t & 0\end{bmatrix}^{T}\eeq 
Here, we have used Eq.\ref{aastar} and $T_\text{DMVP}$ is just the transfer matrix through a DMVP barrier. 
We can again interpret $M_{\phi}^{-1}M_{\theta_1}$ as the phase shift at the first boundary and $M_{\theta_2}^{-1} M_{\phi}$ as the phase shift at the last boundary. These two boundaries are special since these are the interfaces of the magnetic medium with the non-magnetic region. What appears in the middle is the transmission through DMVP barriers repeated $n$ times. A representative plot is given in Fig.\ref{fig6}.
As can be seen, a practical Bragg reflector with a high enough reflectance can be realized with just a few periods of DMVP barriers. A Bragg reflector with large $n$ is broadly similar to an infinite periodic lattice. Particularly at low $B$ ($0.1T$) around $\phi=0$, there is high transmission and $R=0$ for both structures. Similarly, at high $B$ ($3T$), there 
is strong supression of transmission near $\phi=0$ in both cases.
\begin{figure}[ht]
\begin{center}
\begin{psfrags}%
\psfragscanon%
%
\psfrag{s01}[t][t]{\color[rgb]{0,0,0}\setlength{\tabcolsep}{0pt}\begin{tabular}{c}n\end{tabular}}%
\psfrag{s02}[b][b]{\color[rgb]{0,0,0}\setlength{\tabcolsep}{0pt}\begin{tabular}{c}$\langle T(B)\rangle$ (arb. units)\end{tabular}}%
%
\psfrag{x01}[t][t]{$0$}%
\psfrag{x02}[t][t]{$0.1$}%
\psfrag{x03}[t][t]{$0.2$}%
\psfrag{x04}[t][t]{$0.3$}%
\psfrag{x05}[t][t]{$0.4$}%
\psfrag{x06}[t][t]{$0.5$}%
\psfrag{x07}[t][t]{$0.6$}%
\psfrag{x08}[t][t]{$0.7$}%
\psfrag{x09}[t][t]{$0.8$}%
\psfrag{x10}[t][t]{$0.9$}%
\psfrag{x11}[t][t]{$1$}%
\psfrag{x12}[t][t]{$1$}%
\psfrag{x13}[t][t]{$2$}%
\psfrag{x14}[t][t]{$3$}%
\psfrag{x15}[t][t]{$4$}%
\psfrag{x16}[t][t]{$5$}%
\psfrag{x17}[t][t]{$6$}%
\psfrag{x18}[t][t]{$7$}%
\psfrag{x19}[t][t]{$8$}%
\psfrag{x20}[t][t]{$9$}%
\psfrag{x21}[t][t]{$10$}%
%
\psfrag{v01}[r][r]{$0$}%
\psfrag{v02}[r][r]{$0.1$}%
\psfrag{v03}[r][r]{$0.2$}%
\psfrag{v04}[r][r]{$0.3$}%
\psfrag{v05}[r][r]{$0.4$}%
\psfrag{v06}[r][r]{$0.5$}%
\psfrag{v07}[r][r]{$0.6$}%
\psfrag{v08}[r][r]{$0.7$}%
\psfrag{v09}[r][r]{$0.8$}%
\psfrag{v10}[r][r]{$0.9$}%
\psfrag{v11}[r][r]{$1$}%
\psfrag{v12}[r][r]{$0$}%
\psfrag{v13}[r][r]{$0.5$}%
\psfrag{v14}[r][r]{$1$}%
\psfrag{v15}[r][r]{$1.5$}%
\psfrag{v16}[r][r]{$2$}%
%
\resizebox{7.2cm}{!}{\includegraphics{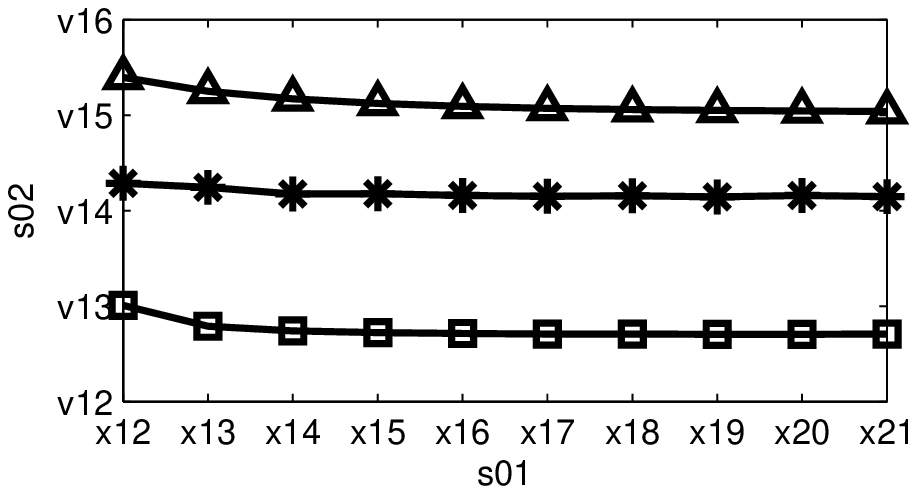}}%
\end{psfrags}%
\end{center}
\caption{\em
Variation of current $\langle T(B)\rangle$ through a Bragg reflector with period n. $B$=$0.1$T ($\triangle$),  $1$T ($\ast$), $3$T ($\Box$) and $d$=$100$nm.}
\label{fig7}
\end{figure}

The higher reflectance of a Bragg reflector will strongly suppress transport. We define the average transmission as  transmittance $T(\phi)$ multiplied by the $x$-component of velocity integrated over all angles of incidence for a given $B$ and $d$ such that  
 \beq \langle T(B)\rangle = 2v_{F}\int_{0}^{\frac{\pi}{2}}d\phi \cos \phi T(\phi) \label{tcurrent} \eeq
Fig.\ref{fig7} plots $\langle T(B)\rangle$ for various $B$ and $d$ and shows transmission is strongly suppressed with increased magnetic field and this  happens within a few periods of the Bragg reflector.  The above formula, when generalized for a range of energy levels, leads to the conductance of the structure, which could be measured experimentally.

\begin{figure}[ht]
\begin{center}
\resizebox{!}{4cm}{\includegraphics{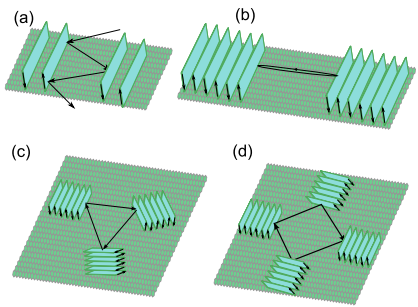}}%
\end{center}
\caption{{\em Bragg reflectors with MVP barriers used as a magnetic waveguide (a), and as resonant cavities (b-d).}}
\label{fig3}
\end{figure}

We have thus shown that reflectance can be controlled by suitably modifying the strength and locations of the magnetic barriers and thereby changing the refractive index of the intervening medium in a novel manner. This principle could be the basis of more elaborate structures depicted in Fig.\ref{fig3}. In a magnetic waveguide (Fig.\ref{fig3}(a)), reflection must be high at desired propagation angles, which could be manipulated by changing the magnetic field. For the resonant cavity shown in Fig.\ref{fig3}(b), high reflection is needed near normal incidence. Geometries such as three-mirror or four-mirror cavities could be used for high reflection at other angles.

To conclude, we have shown that transport through MVP barriers can be understood in terms of propagation of light through periodic stratified media. This analogy can partially be attributed to the fact that Eq.\ref{momentum} describing Dirac fermions and the Maxwell equations \cite{berreman} are both linear wave equations, although there are some key differences that have been pointed out in this paper. 

This formalism describes transport in the ballistic regime, which corresponds to the case of pristine, low-doped Graphene. The same treatment can be also extended to non-relativistic electrons. This picture will be modified when the  effects of disorder and electron-electron interactions are included. 

Using these concepts, practical devices such as Bragg reflectors for manipulating Dirac electrons in graphene can be made.  Such barriers suppress Klein tunneling, thereby achieving confinement in graphene which can be seen through strong supression of transmission of electrons.

\acknowledgments 
We thank G. Baskaran, V. Fal'ko, C.-H. Park and F.M. Peeters for useful comments. The authors acknowledge financial support by IRD Unit, IIT Delhi.

 

\begin{thebibliography}{99}
 \bibitem{SIVAN90}U. Sivan {\it et al}, Phys. Rev. B {\bf 41}, 7937 (1990); S. Datta, 
{\it Electronic Transport in Mesoscopic Systems}, {\bf Chapter 7}, Cambridge University Press,(2005).
\bibitem{GeimreviewKSN1KSN2YZ1Castroneto}A. Geim and K. S. Novoselov, Nat. Mater., {\bf 6}, 183 (2007); K. S. Novoselov {\it et al}, Science, {\bf 306}, 666, (2004); K. S. Novoselov {\it et al}, Nature, {\bf 438}, 197 (2005); Y. Zhang {\it et al}, Nature, {\bf 438}, 201 (2005); A. H. Castro Neto {\it et al}, Rev. Mod. Phys., {\bf 81}, 109 (2009).
\bibitem{Falko}V.V. Cheianov, V. Fal'ko and B.L. Altshuler, Science, {\bf 315}, 1252 (2007).
\bibitem{Veselago}V.G. Veselago, Sov. Phys. Usp., {\bf 10}, 509 (1968); 
J.B. Pendry, Nature, {\bf 423}, 22 (2003).
\bibitem{KSN3}K. S. Novoselov {\it et al}, Nat. Phys., {\bf 2}, 620 (2006).
\bibitem{Beenakker}C. W. J. Beenakker, Rev. Mod. Phys., {\bf 80}, 1337 (2008).
\bibitem{MPV94}A. Matulis, F. M. Peeters and P. Vasilopoulos, Phys. Rev. Lett, {\bf 72}, 1518 (1994);
I. S. Ibrahim and F. M. Peeters, Am. J. Phys. {\bf 63}, 171 (1995).
\bibitem{PG95}J. K. You, L. Zhang, P. K. Ghosh, Phys. Rev. B. {\bf 52}, 17243 (1995).
\bibitem{ZC08}F. Zhai and K. Chang, Phys. Rev. B, {\bf 77}, 113409 (2008).
\bibitem{MDE07} A. De Martino, L. Dell'Anna and R. Egger, Phys. Rev. Lett, {\bf 98}, 066802 (2007).
\bibitem{masir} M.R. Masir {\it et al}, Phys. Rev. B, {\bf 77}, 235443 (2008).
\bibitem{terriskrawczykbader} B.D. Terris and T Thomson, J. Phys. D: Appl. Phys. {\bf 38} R199 (2005); M. Krawczyk and H. Puszkarski, Phys. Rev. B, {\bf 77} 054437 (2008); S.D. Bader, Rev. Mod. Phys., {\bf 78}, 1 (2006).
\bibitem{sunsci}S.H. Sun {\it et al}, Science {\bf 287}, 1989 (2000).
\bibitem{zhengFePt}H. Zheng {\it et al}, Appl. Phys. Lett. {\bf 80}, 2583 (2002).
\bibitem{bergmann}K. von Bergmann {\it et al}, Phys. Rev. Lett. {\bf 96}, 167203 (2006).
\bibitem{Ando}N.H. Shon and T. Ando, J. Phys. Soc. Jpn., {\bf 67}, 2421 (1998).
\bibitem{MSSG2} M. Sharma and S. Ghosh (in preparation).
\bibitem{GhoshOroszlany} L. Oroszl\`{a}ny {\it et al}, Phys. Rev. B {\bf 77}, 081403 (R) (2008); T. K. Ghosh {\it et al}, Phys. Rev. B {\bf 77}, 081404(R) (2008).
\bibitem{yariv1}P. Yeh, A. Yariv and C.S. Hong, J. Opt. Soc. Am., {\bf 67}, 423 (1977).
\bibitem{periodic}B. H. J. Mckellar amd G. J. Stephenson, Jr., Phys. Rev. C {\bf 35}, 2262 (1987);
B. Mendez {\it et al}, J. Phys. A: Math. Gen. {\bf 26}, 171 (1993); P. Strange, {\it Relativistic Quantum Mechanics}, Cambridge University Press, (1998);V.V. Cheianov and V. Fal'ko, Phys. Rev. B, {\bf 74}, 041403 (2006). M. Barbier {\it et al}, Phys. Rev. B {\bf 77}, 115446 (2008); C.-H. Park {\it et al}, Nature Phys., {\bf 4}, 213 (2008).
\bibitem{berreman} D.W. Berreman, J. Opt. Soc. Am., {\bf 62}, 502 (1972).
\end{thebibliography}
\end{document}